\documentclass{jfm}

\usepackage{graphicx}
\usepackage{dcolumn}

\usepackage{amsmath,mathtools}
\usepackage{color}
\usepackage{psfrag}

\usepackage{natbib}

\newcommand{\corr}[1]{{#1}}

\newcommand{\la}{\left<}
\newcommand{\ra}{\right>}

\shorttitle{Transition to the ultimate regime}
\shortauthor{V. Bouillaut, S. Lepot, S. Auma\^itre, B. Gallet}
\title{Transition to the ultimate regime in a radiatively driven convection experiment}

\author{Vincent Bouillaut, Simon Lepot, S\'ebastien Auma\^itre, Basile Gallet}

\affiliation{Service de Physique de l'\'Etat Condens\'e, CEA,
CNRS UMR 3680, Universit\'e Paris-Saclay, CEA Saclay, 91191 Gif-sur-Yvette,
France}

\date{\today}

\begin{document}

\maketitle

\begin{abstract}
We report on the transition between two regimes of heat transport in a radiatively driven convection experiment, where a fluid gets heated up within a tunable heating length $\ell$ in the vicinity of the bottom of the tank. The first regime is similar to the one observed in standard Rayleigh-B\'enard experiments, the Nusselt number $Nu$ being related to the Rayleigh number $Ra$ through the power-law $Nu \sim Ra^{1/3}$. The second regime corresponds to the ``ultimate'' or mixing-length scaling regime of thermal convection, where $Nu$ varies as the square-root of $Ra$. Evidence for these two scaling regimes have been reported in Lepot et al. (Proc. Nat. Acad. Sci. U S A, {\bf 115}, 36, 2018), and we now study in detail how the system transitions from one to the other. We propose a simple model describing radiatively driven convection in the mixing-length regime. \corr{It leads to the scaling relation $Nu \sim \frac{\ell}{H} Pr^{1/2} Ra^{1/2}$,} where $H$ is the height of the cell, thereby allowing us to deduce the values of $Ra$ and $Nu$ at which the system transitions from one regime to the other. These predictions are confirmed by the experimental data gathered at various $Ra$ and $\ell$. \corr{We conclude by showing that boundary layer corrections can persistently modify the Prandtl number dependence of $Nu$ at large $Ra$, for $Pr \gtrsim 1$.}
\end{abstract}

\section{Introduction} 

In many geophysical and astrophysical flows, turbulent convection is driven by local internal heating. For instance, the absorption of sunlight within the first few meters of water inside frozen lakes induces convective mixing and penetrative convection \citep{Farmer,Bengtsson,Jonas,Mironov,Ulloa,Lecoanet,Toppaladoddi}. A second example is the interior of stars, where, depending on the stellar mass, internal heating due to thermonuclear reactions can directly overlap with the convective region \citep{Kippenhahn,Barker}. Inside the Earth's mantle, radioactive decay induces internal heating as well \citep{Davaille,Limare}. Finally, convection driven by a flux of neutrinos within collapsing stellar cores is believed to affect the shape of supernovae explosions \citep{Herant,Janka,Radice,Kazeroni}. To reproduce such convection in the laboratory, in a previous study we introduced an experimental setup in which a turbulent flow is driven by the absorption of a flux of light \citep{Lepot}: A powerful spotlight shines from below at an experimental cell with a transparent bottom plate. The cell contains a mixture of water and dye, which absorbs the light flux and converts it into heat. The heating is therefore localized near the bottom of the tank, on a typical height $\ell$ that can be tuned through the concentration of the dye.
We showed that when $\ell$ is much smaller than the boundary layers near the bottom of the tank, radiative heating is similar to that of a standard Rayleigh-B\'enard (RB) experiment, i.e., of a plate heated at constant power (through a Boussinesq symmetry, the system is equivalent  to convection driven by uniform internal heating together with a fixed-flux cooling upper boundary, as introduced by \citet{Goluskin2015}). In terms of Nusselt and Rayleigh numbers $Nu$ and $Ra$ (see (\ref{defRaNu}) for definitions), we measured a power-law close to $Nu \sim Ra^{1/3}$, which corresponds to the regime of standard RB experiments \citep{Malkus,Chavanne97,Niemela,Chavanne01,Alhers,Roche}\corr{: the heat transport efficiency is restricted by the diffusion of heat across the marginally stable boundary layer located near the bottom plate. The thickness of this marginally stable boundary layer is independent of the height of the fluid layer, and so is the relation between the heat flux and the temperature drop across the cell, hence the scaling-law $Nu \sim Ra^{1/3}$.} More interestingly, when $\ell$ is large enough for heat to be input directly into the bulk turbulent flow, we observed that radiative heating leads to the mixing-length or ``ultimate'' scaling regime, $Nu \sim Ra^{1/2}$, which corresponds to a fully turbulent regime where the molecular diffusion coefficients are irrelevant \citep{Spiegel63,Kraichnan,Spiegel}.

The goal of the present study is to understand the transition between these two regimes: what happens for intermediate values of the heating length $\ell$? Indeed, as compared to RB studies, our setup has an additional dimensionless parameter: the dimensionless absorption length $\ell/H$, where $H$ denotes the height of the fluid layer. What is the dependence of the Nusselt number on this new parameter? Dimensional analysis leads to:
\begin{equation}
Nu={\cal F}(\ell/H,Ra,Pr) \, , \label{Nufunctional}
\end{equation}
and because $\ell/H$ governs the transition between two different scaling-regimes, the relation (\ref{Nufunctional}) does not in general take the form of a power-law. However, once we are in a given scaling regime we can write (\ref{Nufunctional}) as a power-law:
\begin{equation}
Nu = \text{const.} \left( \frac{\ell}{H} \right)^\beta Ra^\gamma Pr^\chi \, . \label{scalinggeneral}
\end{equation}
In the following we propose simple models leading to predictions for the values of the exponents $\beta$, $\gamma$ and $\chi$, before confronting these predictions to the experimental data.

We introduce the experimental setup in section \ref{sec:expsetup}. In section \ref{sec:transition}, we present the experimental data for the Nusselt number, before introducing a simple model that predicts the scaling behavior (\ref{scalinggeneral}) in the ultimate regime, with the exponents $\beta=1$, $\gamma=1/2$ and $\chi=1/2$. We show that the experimental data are compatible with these values of $\beta$ and $\gamma$. The discussion section \ref{sec:discussion} focuses on the dependence in $Pr$, which cannot be probed within the present experimental setup. While the value $\chi=1/2$ should be achieved at low $Pr$, a refinement of the model indicates that, for finite or large $Pr$, the injection of even a tiny fraction of the radiative heat flux into the boundary layers could result in a persistent modification of the exponent $\chi$, while maintaining $\beta=1$ and $\gamma=1/2$.

\section{Experimental setup \label{sec:expsetup}}

\begin{figure}
   \centering{\includegraphics[width=7cm]{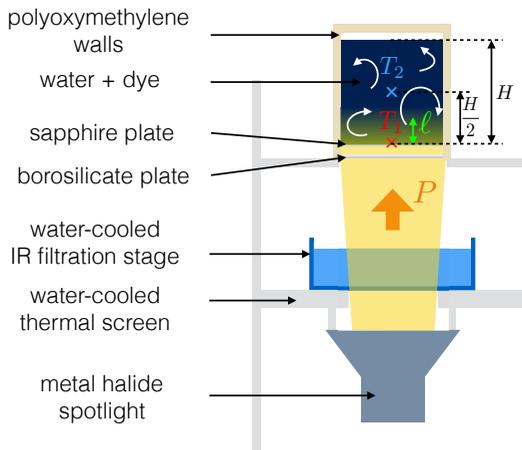}}   
    \caption{{\bf Radiatively driven convection in the laboratory:} A powerful spotlight shines from below at a cell containing a mixture of water and dye. This triggers volumic heating near the bottom plate, over a typical length $\ell$ that can be tuned through the concentration of the dye. \label{fig:setup}}
\end{figure}

\subsection{Radiative heating in the laboratory}

The experimental setup is sketched in figure \ref{fig:setup}. It has been described in a previous publication \citep{Lepot} and we only mention its key characteristics here. A 2500W metal-halide spotlight shines at a cylindrical experimental cell of radius $R=10\,$cm containing a homogeneous mixture of water and carbon black dye. The sidewalls of the tank are made of polyoxymethylene, while the bottom boundary is a transparent sapphire plate. The light flux penetrates into the tank, where it is absorbed by the dye and turned into heat. Beer-Lambert law states that the light flux inside the tank then decreases exponentially with the height $z$ measured upwards from the bottom plate, and so does the heating rate. The bulk heating rate $Q(z)$ inside the tank therefore reads:
\begin{equation}
Q(z)=\frac{P}{\ell} \exp(-z/\ell) \, ,
\end{equation}
where $P$ is the heat flux radiated by the spotlight in the form of visible light (in units of W.m$^{-2}$). The absorption length $\ell$ is inversely proportional to the dye concentration. By changing the latter, we can achieve either standard RB heating, when $\ell$ is much smaller than the boundary layer thickness, or significant heating of the bulk turbulent flow, when $\ell$ is much greater than the boundary layer thickness. 

\subsection{``Secular'' cooling}

A key aspect of the experiment is to avoid boundary layers at the cooling side as well. Indeed, a fixed temperature cooling plate would produce standard boundary layers restricting the heat flux. Because of this cold boundary layer, traditional studies of internally heated convection have led to scaling-laws similar to that of standard RB convection \citep{Kulacki,Goluskin}. We follow a different approach, inspired by the ``secular heating'' invoked in many studies of convection in the Earth interior \citep{Gubbins,Aubert,Landeau}: if we do not cool down the system, the temperature at any point within the fluid drifts with time at a constant rate. On top of this linear drift, the flow develops some stationary internal temperature gradients. If $T({\bf x},t)$ denotes the temperature field inside the tank and $\overline{T}(t)$ its spatial average, one can show easily that $\overline{T}(t)$ increases linearly in time at a rate proportional to the radiative flux of the spotlight:
\begin{equation}
\frac{\mathrm{d}\overline{T}}{\mathrm{d}t} = \frac{P}{\rho C H} \left( 1-e^{-H/\ell}\right)\, , \label{eqslope}
\end{equation}
where $\rho$ is the average density of the fluid and $C$ its specific heat capacity. The local deviation from the mean temperature is $\theta({\bf x},t)=T({\bf x},t)-\overline{T}(t)$. One can easily show that the field $\theta({\bf x},t)$ obeys the equations of Boussinesq convection for a fluid that is radiatively heated and cooled uniformly in space. In particular, the heat equation becomes:
\begin{equation}
\partial_t \theta + {\bf u} \cdot {\boldsymbol{\nabla} \theta} =  \kappa \boldsymbol{\nabla}^2 \theta +\frac{1}{\rho C} \left[ Q(z)-\frac{P}{ H}  \left( 1-e^{-H/\ell}\right) \right] \, , \label{eqtheta}
\end{equation}
where $\kappa$ denotes the thermal diffusivity. On average over space, the uniform cooling term -- the second term in the square bracket -- balances the radiative heating rate, so that after a transient $\theta({\bf x},t)$ reaches a statistically steady state. 

\subsection{Measurements and control parameters}

\begin{figure}
   \centering{\includegraphics[width=9cm]{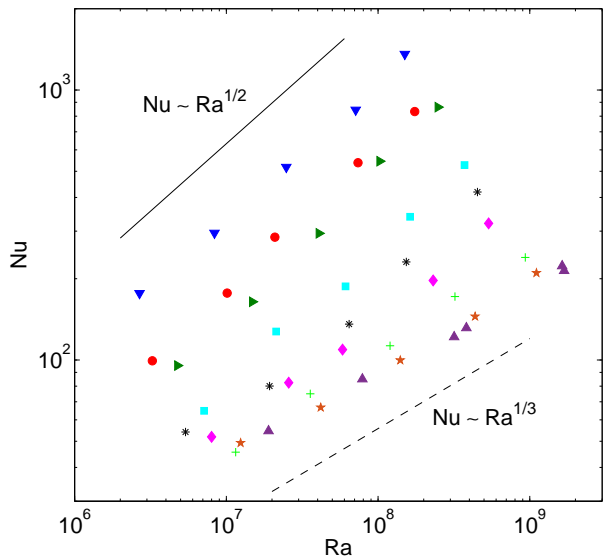}}   
    \caption{Nusselt number as a function of the Rayleigh number for various values of the absorption length $\ell$. At fixed $Ra$, the Nusselt number increases with $\ell/H$. Symbols are
     $\triangle$: $\ell=5 . 10^{-6}$ m; $\star$: $\ell/H=0.0015$;  $+$: $\ell/H=0.0030$;   $\diamond$: $\ell/H=0.0060$;   $*$: $\ell/H=0.012$;  $\square$: $\ell/H=0.024$;   $\triangleright$: $\ell/H=0.048$;    $\circ$: $\ell/H=0.05$;   $\triangledown$: $\ell/H=0.096$. The solid and dashed lines are eyeguides.  \label{fig:NuvsRa}}
\end{figure}

We measure the internal temperature gradients using two thermocouples. The first one touches the bottom sapphire plate and gives access to its temperature $T_1$, while the second one measures the temperature $T_2$ at mid-depth inside the tank. Both probes are centered horizontally. As discussed in the previous subsection, the measured temperature difference $\Delta T=T_1-T_2=\theta_1-\theta_2$ is governed by the Boussinesq equations subject to both radiative heating and uniform ``secular'' cooling.

Metal halide spotlights cannot be operated over a large range of power. To scan a broad range of Rayleigh numbers, we therefore vary the depth $H$ of the fluid layer from $4$ cm to $19$ cm. The second control parameter of the experiment is the dye concentration, which allows us to vary the dimensionless absorption length $\ell/H$ over several orders of magnitude.

A typical experimental run consists in starting with the mixture of water and dye around $8^o$C before turning the spotlight on.  Both temperatures increase with time, \corr{and a stationary temperature difference between the two probes is achieved after a few turnover times (roughly $200$ s).} We keep the part of the temperature signals corresponding to a bottom temperature between $\pm 2^\circ$C of room temperature. We average $\Delta T$ over this time interval, and we extract the heat flux $P$ from the slope of the common temporal drift of the two signals (see equation (\ref{eqslope})). We finally compute the Rayleigh and Nusselt numbers as:
\begin{equation}
Ra=\frac{\alpha g \la \Delta  T \ra H^3}{\kappa \nu} \qquad Nu=\frac{PH}{\lambda \la \Delta T \ra} \, , \label{defRaNu}
\end{equation}
where $\alpha$ denotes the thermal expansion coefficient, $g$ is gravity, $\nu$ is the kinematic viscosity, $\lambda$ is the thermal conductivity, and $\la \cdot \ra$ denotes time average.

\section{From the Rayleigh-B\'enard regime to the mixing-length one \label{sec:transition}}

We have performed several sets of experimental runs for various quantities of dye, i.e., for various dimensionless absorption lengths $\ell/H$. We show in figure \ref{fig:NuvsRa} the corresponding $Nu$ versus $Ra$ curves. We also reproduce the data from \citet{Lepot}, where the absorption length is either $\ell/H<10^{-4}$, or $\ell/H=0.05$. The former case corresponds to a RB situation and displays a power-law behavior $Nu\sim Ra^{0.31}$, while for the latter case heat is input inside the bulk turbulent flow, which leads to a power-law $Nu\sim Ra^{0.54}$ close to the prediction of Spiegel and Kraichnan. The new data points span the transition region between these two limiting regimes. While the curves for the lowest values of $\ell/H$ are superimposed onto the RB case, for larger $\ell/H$ and fixed $Ra$ the Nusselt number increases with $\ell/H$. \corr{The various $Nu$ versus $Ra$ curves of figure  \ref{fig:NuvsRa} are roughly compatible with power-laws. However, while the corresponding power-law fits are very good for extreme values of $\ell/H$, the residuals are larger for intermediate values of $\ell/H$: for instance, the $Nu$ versus $Ra$ curve for $\ell/H=0.012$ exhibits a slight positive  convexity in log scales, which we will argue is a signature of the crossover region between the RB and the ultimate scaling regimes.}
Roughly speaking, the transition to the ultimate regime takes place when radiative heating bypasses the boundary layers and injects the heat directly into the bulk flow, i.e., when $\ell$ is much larger than the boundary layer thickness. This can be achieved either by increasing $\ell/H$ for fixed $Ra$, or by increasing $Ra$ with fixed $\ell/H$ to decrease the boundary layer thickness. In the following we propose a simple model to further investigate this transition.

\begin{figure}
   \centering{\includegraphics[width=9cm]{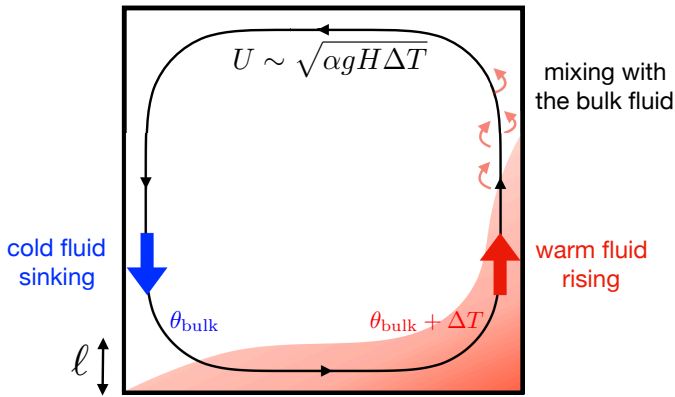}}   
    \caption{\textbf{Model of radiatively heated convection roll:} The black line is a streamline of the mean flow. Near the bottom left corner, a cold fluid element at the bulk temperature enters the heating region. It gets heated as it travels along the bottom boundary, gaining an overall temperature increment of the order of $\Delta T$ during a time of flight $H/U$. The fluid element has maximum temperature as it passes near the bottom-right corner. It then starts rising and follows the mean cellular motion while mixing with the bulk fluid.    \label{fig:roll}}
\end{figure}

\subsection{A roll model}

In figure  \ref{fig:roll} we sketch a simple model to estimate the temperature difference $\Delta T$ within the experimental cell. At large scale, turbulent convective flows typically consist of cellular motion, as sketched in figure \ref{fig:roll}. The typical temperature difference $\Delta T$ inside the cell can be estimated by considering a fluid element evolving on a \corr{streamline} near the periphery of the convective roll. Near the left-hand boundary of the domain in figure \ref{fig:roll}, the fluid particle is close to the bulk temperature. It gets advected downward by the convective roll and enters the heating region. This Lagrangian fluid element then travels close to the bottom boundary, within the heating region. During this phase it gets heated up, its temperature increasing from the bulk temperature $\theta_{\text{bulk}}$ to approximately $\theta_{\text{bulk}}+\Delta T$ as it reaches the bottom-right corner. As long as the particle remains close to the bottom boundary, we have $z \ll \ell \ll H$, and the dominant balance in equation (\ref{eqtheta}) written for the fluid particle reads:
\begin{equation}
\frac{D \theta}{D t} \simeq \frac{1}{\rho C} \left[ Q(z)-\frac{P}{ H}  \left( 1-e^{-H/\ell}\right) \right] \simeq \frac{P}{\rho C \ell} \, ,
\end{equation}
where $\frac{D \cdot}{Dt}$ denotes the total derivative. For a convective roll of unit aspect ratio, the travel time of the fluid element from the bottom-left to the bottom-right corner is $\Delta t \sim H/U$, where $U$ is the typical velocity of the convective roll. Assuming that $U$ follows the free-fall scaling-law:
\begin{equation}
U \sim \sqrt{\alpha g \Delta T H} \, , \label{ffvelocity}
\end{equation}
the temperature increase during the heating phase is estimated as:
\begin{equation}
\Delta T \sim  \frac{P}{\rho C \ell} \Delta t \sim  \frac{P H}{\rho C \ell U} \sim  \frac{P H}{\rho C \ell \sqrt{\alpha g \Delta T H}} \, , \label{DeltaTroll}
\end{equation}
which, in terms of dimensionless quantities, yields:
\begin{equation}
Nu \sim \frac{\ell}{H} Pr^{1/2} Ra^{1/2} \, . \label{ultimatescaling}
\end{equation}
The warm fluid element then starts rising. It exits the heating region and gradually mixes with the bulk fluid as it moves around the cell. It has relaxed to the bulk fluid temperature when it reaches the bottom left corner of the convection roll again, and a new cycle starts.

\subsection{Transition point and rescaling of the data}

\begin{figure}
   \centering{\includegraphics[width=12cm]{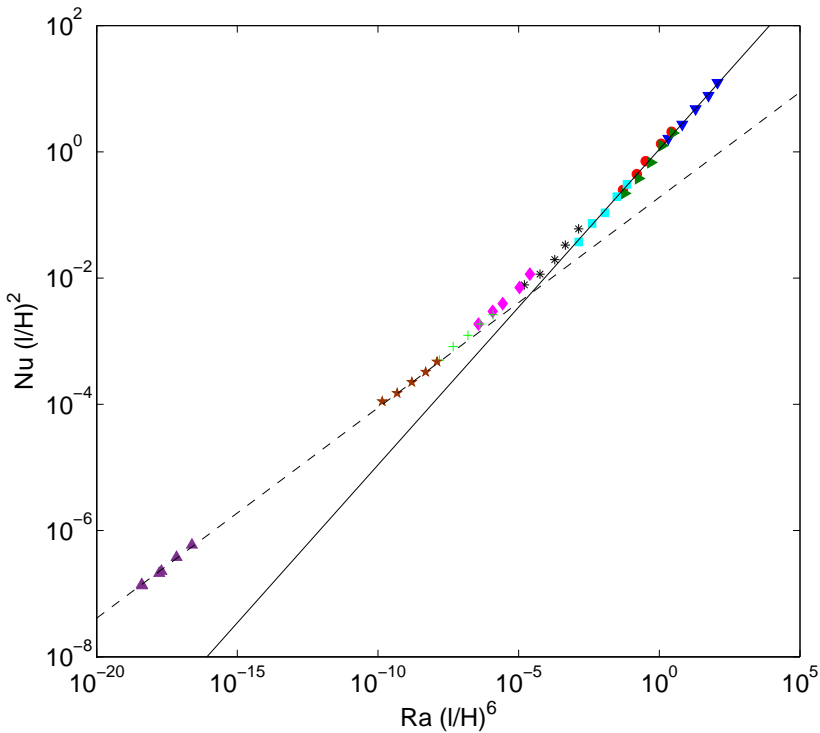}}   
    \caption{Rescaled Nusselt number as a function of the rescaled Rayleigh number, for various values of the absorption length $\ell$ (same symbols as figure \ref{fig:NuvsRa}). The data indicate a clear transition from an exponent $\gamma=1/3$ (dashed-line) to an exponent $\gamma=1/2$ (solid line).\label{fig:NuvsRarescaled}}
\end{figure}

To test the compatibility between the prediction (\ref{ultimatescaling}) and the experimental data, one can focus on the transition between the two asymptotic regimes. For small absorption length or small Rayleigh number, we expect to recover the scaling regime of Rayleigh-B\'enard convection. A marginally stable boundary layer argument then yields the power-law $Nu\sim Ra^{1/3}$, i.e., $\gamma=1/3$ and $\beta=\chi=0$ in the general scaling relation (\ref{scalinggeneral}). For higher Rayleigh numbers, $\ell$ is much larger than the boundary layer thickness. Heat is input predominantly inside the bulk turbulent flow and the regime (\ref{ultimatescaling}) eventually sets in, with $\beta=1$ and $\gamma=\chi=1/2$. \corr{As $Ra$ increases from low values, the RB regime should hold until the thickness $\delta$ of the marginally stable thermal boundary layer becomes comparable to $\ell$. Indeed, a similar argument for convection over rough plates successfully predicts a departure from the standard RB regime when $\delta$ is comparable to the typical roughness height \citep{Shen,Toppaladoddi2017,Xie,Rusaouen}. However, in the present setup the transition is slightly more subtle, and $\ell \sim \delta$ is not the threshold where the scaling-law (\ref{ultimatescaling}) sets in. To see this, one can perform an energy budget inside the heating region $z\lesssim \ell$ in figure \ref{fig:roll}: fluid enters this region near the bottom left corner at temperature $\theta_{\text{bulk}}$ and exits the domain near the bottom-right corner, with a temperature $\theta_{\text{bulk}}+\Delta T$. The power (heat per unit time, in Joules per second) evacuated from this region by the large-scale roll is therefore $\phi_U \sim H \ell U \rho C  \Delta T $, while the power input by the radiative heating is $P H^2$. If we substitute the optimistic free-fall estimate (\ref{ffvelocity}) for $U$, the ratio of the former over the latter becomes:}
\begin{equation}
\frac{\phi_U}{P H^2} \sim \frac{\ell}{H} \frac{Pr^{1/2} Ra^{1/2}}{Nu} \, . \label{phiUoverQinj}
\end{equation}
\corr{At the point where $\ell\sim\delta$, the RB scaling still holds: substituting $Nu\sim Ra^{1/3}$ and $\ell \sim \delta \sim Ra^{-1/3}$ into (\ref{phiUoverQinj}) yields ${\phi_U}/{P H^2}\sim Pr^{1/2} Ra^{-1/6} \ll 1$. We conclude that the roll is too slow to efficiently extract the heat input radiatively inside the heating region when $\delta=\ell$. The roll mechanism described above therefore sets in at higher Rayleigh numbers. The ratio (\ref{phiUoverQinj}) is then of the order of unity, which again yields the scaling-law (\ref{ultimatescaling}).}
\corr{Because of the limited power of the spotlight, these two transitions -- the end of the RB scaling regime and the beginning of the ultimate one -- cannot be distinguished in our experiment. Instead, we will show that the data is well described by a single overall transition point $(Ra_\text{tr},Nu_\text{tr})$ lying at the intersection between the two extreme scaling-laws $Nu\sim Ra^{1/3}$ and (\ref{ultimatescaling}):} 
\begin{equation}
Nu_\text{tr} \sim Ra_\text{tr}^{1/3} \sim \frac{\ell}{H} Pr^{1/2} Ra_\text{tr}^{1/2} \, ,
\end{equation}
from which we deduce:
\begin{equation}
Ra_\text{tr} \sim Pr^{-3} \left( \frac{\ell}{H}\right)^{-6} \qquad  Nu_\text{tr} \sim Pr^{-1} \left( \frac{\ell}{H}\right)^{-2}    \, . \label{predictedtrvalues}
\end{equation}
One way to test the predictions of this model is to plot the Rayleigh and Nusselt numbers rescaled by their values at the transition, i.e., $Nu/Nu_\text{tr}$ as a function of $Ra/Ra_\text{tr}$. In figure \ref{fig:NuvsRarescaled}, we thus plot $Nu \, (\ell/H)^2$ as a function of $Ra \, (\ell/H)^6$. In this representation the data obtained for various values of the absorption length $\ell$ collapse onto a single master curve. The latter starts off with an exponent $1/3$, before transiting to a second power-law with an exponent compatible with the $1/2$ prediction of the model above. This representation confirms the dependence of $Nu$ with $\ell/H$ and $Ra$ in the two regimes.

\section{Discussion: dependence in $Pr$ and persistent boundary layers \label{sec:discussion}}

While the roll model described above successfully predicts the dependence of the Nusselt number with $\ell/H$ and $Ra$ in the ultimate regime, the predicted dependence with $Pr$ cannot be tested with the present experimental setup. As a word of caution, we therefore wish to discuss how the boundary layers can affect the $Pr$-dependence of the Nusselt number. Coming back to the general scaling relation (\ref{scalinggeneral}), we will show that these boundary layers can induce a persistent modification of the exponent $\chi$ at high Rayleigh numbers, while leaving the values $\beta=1$ and $\gamma=1/2$ unchanged.

\begin{figure}
   \centering{\includegraphics[width=7cm]{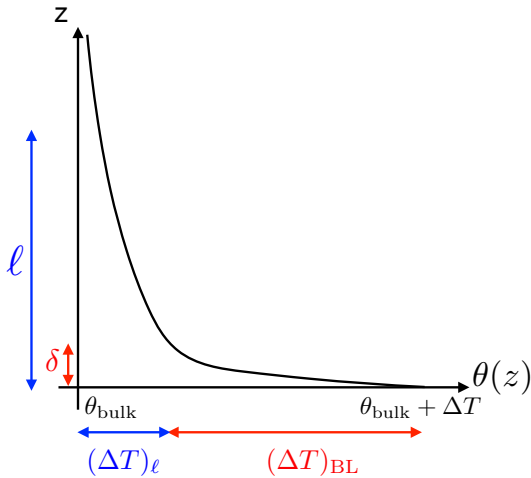}}   
    \caption{Schematic vertical temperature profile showing the temperature drop $(\Delta T)_\ell$ within the absorption region, together with an additional temperature drop $(\Delta T)_\text{BL}$ within a boundary layer of thickness $\delta \ll \ell$.   \label{fig:BL}}
\end{figure}

Near the bottom wall is a boundary layer, for the velocity field to satisfy the no-slip boundary condition. The velocity is reduced within this boundary layer ; coming back to the simple picture of figure \ref{fig:roll}, the heating phase is longer if the fluid element travels on a streamline that is contained inside the boundary layer. It accumulates more heat than fluid particles travelling outside the boundary layer, and reaches higher temperatures. There is therefore also a boundary layer for the temperature field. A schematic of the resulting horizontally averaged and time averaged temperature profile is provided in figure \ref{fig:BL}. We expect a temperature drop $(\Delta T)_\ell$ on a vertical scale $\ell$, associated to the roll model described above, together with an additional temperature difference $(\Delta T)_\text{BL}$ associated to the boundary layer region, within which diffusion plays a central role. 
Inside a boundary layer of thickness $\delta \ll \ell$, the heat input by the radiative forcing is approximately $Q(z=0) \times \, \delta = P\, \times \delta / \ell$. This heat flux is diffused outside of the boundary layer, which leads to:
\begin{equation}
P \frac{\delta}{\ell} \sim \frac{\lambda (\Delta T)_\text{BL}}{\delta} \, .
\end{equation}
From this equality we deduce $(\Delta T)_\text{BL}$ in terms of thickness $\delta$ of the temperature boundary layer. The total temperature drop $\Delta T$ is then the sum of $(\Delta T)_\text{BL}$ and of the temperature drop $(\Delta T)_\ell$ outside of the boundary layer, which we estimate using expression (\ref{DeltaTroll}). We obtain:
\begin{equation}
\Delta T = (\Delta T)_\ell + (\Delta T)_\text{BL} = c_0 \frac{P H^2}{\lambda \ell} Pr^{-1/2} Ra^{-1/2} + c_1 \frac{P H^2}{\lambda \ell} \times \frac{\delta^2}{H^2} \, , \label{DeltaTtot}
\end{equation}
where the $(c_i)_{i \in \mathbb{N}}$ are dimensionless constants. 
The next step is to insert scaling-laws for the boundary-layer thickness $\delta$, to examine their consequences on the scaling relation (\ref{scalinggeneral}). We distinguish between low- and large-Prandtl-number fluids.

\subsection{Low-Prandtl-number fluids}

Let us denote as $\delta_\nu$ the thickness of the velocity boundary layer. The standard estimate for $\delta_\nu$ is:
\begin{equation}
\delta_\nu \sim \frac{H}{\sqrt{Re}} \, ,
\end{equation}
where the Reynolds number is defined as $Re=U H/\nu$. Substituting the free-fall velocity estimate (\ref{ffvelocity}) for $U$ yields:
\begin{equation}
\delta_\nu \sim H \, Pr^{1/4} Ra^{-1/4} \, . \label{scalingdeltanu}
\end{equation}
In a low-Prandtl-number fluid, the temperature field shares this boundary layer thickness, as it gets mixed very efficiently by the turbulent flow outside of it. Inserting $\delta=\delta_\nu$ into expression (\ref{DeltaTtot}) leads to:
\begin{equation}
\Delta T = c_0 \frac{P H^2}{\lambda \ell} Pr^{-1/2} Ra^{-1/2} + c_2 \frac{P H^2}{\lambda \ell} Pr^{1/2} Ra^{-1/2} \, . \label{DeltaTtotlowPr}
\end{equation}
The boundary layer correction to the temperature drop -- the second term in (\ref{DeltaTtotlowPr}) -- is smaller than the main contribution  of the roll model by a factor $Pr$. Although it may be possible to detect it for moderately low $Pr$, it is negligible for $Pr \ll 1$.


\subsection{Large-Prandtl-number fluids}

If the Prandtl number is much greater than unity, the boundary layer of the temperature field is much thinner than $\delta_\nu$: the temperature drop associated to the thermal boundary layer takes place within the velocity boundary layer. The velocity field in this region can be approximated by a uniform shear flow, the shear being $S \sim U/\delta_{\nu}$. Following \citet{Shraiman}, the thermal boundary layer thickness $\delta$ is then:
\begin{equation}
\delta \sim \left( \frac{\kappa H}{S} \right)^{1/3} \sim \left( \frac{\kappa H \delta_\nu}{U} \right)^{1/3} \sim H Pr^{-1/12} Ra^{-1/4} \, ,
\end{equation}
where we have substituted the estimates (\ref{ffvelocity}) and (\ref{scalingdeltanu}) for $U$ and $\delta_\nu$. Inserting this expression for the thermal boundary layer thickness into (\ref{DeltaTtot}) yields:
\begin{equation}
\Delta T = c_0 \frac{P H^2}{\lambda \ell} Pr^{-1/2} Ra^{-1/2} + c_3 \frac{P H^2}{\lambda \ell} Pr^{-1/6} Ra^{-1/2} \, , \label{DeltaTtothighPr}
\end{equation}
The boundary layer correction to $\Delta T$ is important in this large-$Pr$ regime, as it becomes the main contribution to $\Delta T$ in the limit $Pr \gg 1$. In this limit, the scaling-law for the Nusselt number (\ref{defRaNu}) becomes:
\begin{equation}
Nu \sim \frac{\ell}{H} Pr^{1/6} Ra^{1/2} \, . \label{scalinghighPr}
\end{equation}
The boundary layer correction leads to $\chi=1/6$ instead of $\chi=1/2$, with still $\beta=1$ and $\gamma=1/2$. This is a persistent modification of $\chi$, in the sense that the scaling-law is modified up to arbitrarily large Rayleigh number. 
While this discussion section is only here to highlight possible modifications of the exponent $\chi$ by boundary layer dynamics, the precise determination of $\chi$ remains an experimental challenge. A dedicated numerical study may be a simpler approach to address the dependence of $Nu$ over several decades of $Pr$.  In the meantime, we shall compare the results of this study to convective flows inside frozen great lakes, the Prandtl number of which is only twice our experimental value. To wit, it is desirable to re-express the transition point between the RB and ultimate regimes in terms of control parameters only, independent of the measured temperature drop $\Delta T$. We thus introduce the flux-based Rayleigh number $Ra_P= Nu\times Ra = \alpha g P H^4 / \lambda \kappa \nu$. On the one hand, equation (\ref{predictedtrvalues}) together with the data of figure \ref{fig:NuvsRarescaled} indicate that the ultimate scaling regime sets in for $Ra_P (\ell/H)^8 \gtrsim 3. 10^{-7}$, for the Prandtl number of water at $28 ^\circ$C. We can compare this criterion to the typical value of $Ra_P$ for frozen great lakes in the spring \citep{Mironov,Ulloa}: with a light flux $P \simeq 100\,$W.m$^{-2}$, an absorption length $\ell \simeq 1\,$m and a mixed-layer depth $H$ ranging from $4\,$m to $40\,$m, we obtain $Ra_P (\ell/H)^8$ in the range $10^5-10^9$, well inside the region of parameter space where the mixing-length scaling regime holds. This confirms that radiative heating -- as opposed to fixed-flux heating at the boundary -- is a key ingredient of any laboratory of numerical setup aimed at describing the thermal structure of such lakes.

\corr{We thank V. Padilla for his help during the development of the experimental setup. This research is supported by the European Research Council under grant agreement FLAVE 757239, and by Labex PALM ANR-10-LABX-0039.}


\end{document}